\documentclass[5p]{elsarticle}
\usepackage{graphicx} 

\begin{document}

\title{Sequences of Gluing Bifurcations in an Analog Electronic Circuit}

\author[alm]{Sayat N.~Akhtanov}
\author[alm]{Zeinulla Zh. Zhanabaev}
\author[mz]{Michael A. Zaks$^*$}
\address[alm]{Physical-Technical Department, 
Al Farabi Kazakh National University, 
Al Farabi Av. 71, Almaty, 050038 Kazakhstan}
\address[mz]{Institute of Mathematics, Humboldt University, 
Rudower Chaussee 25, D-12489 Berlin, Germany}

\begin{abstract}
We report on the experimental investigation of gluing bifurcations 
in the analog electronic circuit which models a dynamical system of 
the third order: Lorenz equations with an additional quadratic nonlinearity. 
Variation of one of the resistances in the circuit changes the coefficient
at this nonlinearity and enables transition from the Lorenz route to chaos
to a different scenario which 
leads, through the sequence of homoclinic bifurcations, from
periodic oscillations of the voltage to the chaotic state.
A single bifurcation ``glues'' in the phase space two stable periodic orbits 
and creates a new one, with the doubled length: 
a bifurcation sequence results in the birth of the chaotic attractor.
\end{abstract}


\maketitle
\noindent $^*$\ \ Corresponding author. Tel.: +49 30 20932317,\\ Fax +49  30 20931842.\\
 E-mail address: zaks@math.hu-berlin.de 
\section{Introduction}
Studies of routes from order to chaos 
in families of low-dimensional dynamical systems 
usually started with theoretical and numerical investigations,
which were soon followed by detailed experiments.
Accurate experimental verification 
not only recovered the qualitative picture of the transition 
but was also able to resolve its quantitative characteristics. 
Universal sequence of period-doubling bifurcations, 
onset of chaos via the breakup of the quasiperiodic oscillations 
as well as scaling laws in different kinds of intermittency 
have been documented in numerous 
mechanical, hydrodynamical, optical, chemical etc. 
experiments (see references to original publications e.g. 
in~\cite{Schuster-88,Argyris-Faust-Haase-94}).  

There is, however, a group of scenarios which is
well understood theoretically, but
has received less attention from experimentalists:
these are the routes to chaos via the sequences of 
so-called {\em gluing bifurcations}. 
Along such routes, pairs of stable periodic orbits 
come close in the phase space, 
recombine and form new stable periodic orbits 
which are more complicated than the original ones. 
Each recombination is mediated by
two trajectories, homoclinic to the same saddle point. 
Homoclinic bifurcations often occur in nonlinear dynamics: 
in the Lorenz equations~\cite{Lorenz-63}, 
the birth of the chaotic set in the phase space owes to the so-called 
``homoclinic explosion''~\cite{Sparrow-82,Afraimovich-Bykov-Shilnikov-77}.
The gluing bifurcation differs from the homoclinic explosion both in
the number of newborn periodic orbits and in their stability.
The explosion generates, in a single act of creation,
the countable set of periodic orbits, each of them asymptotically unstable; 
taken together, they form the kind of ``skeleton'' 
for the emerging chaotic attractor. 
In contrast, a gluing bifurcation, taken alone, 
produces just one or two stable periodic orbits~\cite{Gambaudo_etal_88}. 
However, in the course of the sequence (``scenario'') of such bifurcations the 
shape of the attracting orbit gets more and more involved, and its length 
grows, until the whole development culminates in the chaotic attractor.  
Which of two cases -- a homoclinic explosion or a gluing bifurcation -- 
takes place in the particular family of dynamical systems, 
is entirely determined by the ratio of the two leading eigenvalues 
of the Jacobian computed at the saddle point. 

From the point of view of an experimentalist,  
gluing scenarios have an unpleasant feature:
vulnerability of their basic building blocks.
Unlike periodic orbits, saddle connections in generic dynamical systems 
are structurally unstable. 
Furthermore, a gluing bifurcation assumes coexistence, 
at the same set of parameter values,
of two homoclinic trajectories to the same saddle point. 
Since, generically, formation of the homoclinic trajectory to a saddle
point of a dissipative system is a codimension-one event, every 
gluing bifurcation is a codimension-two phenomenon.
Hence, the accomplishment of the gluing scenario requires 
either the perfect mirror ($Z_2$) symmetry of the system 
(which transforms a homoclinic orbit into another one, so that they
exist in pairs), or the ability to track in the parameter space 
the sequences of codimension-two events. 
This makes gluing bifurcations a difficult object 
for  laboratory studies: they are sensitive both to fluctuations 
and to imperfections of the experimental  setup. 
Most occurrences of gluing scenarios in families of dynamical systems
were reported in theoretical and numerical studies: in the context of 
hydrodynamics~\cite{Lyubimov-Zaks-83,Busse-Kropp-Zaks-92,Rucklidge-94}, 
nematic liquid crystals~\cite{Demeter-Kramer-99,Carbone-Cipparrone-2001} 
and optothermal devices~\cite{Herrero-98}. 
In the experiments, separate gluing bifurcations were identified 
in the Taylor-Couette flow of the viscous fluid 
between two cylinders~\cite{Abshagen-Pfister-Mullin-2001}
and in the Chua oscillators~\cite{Roy_Dana}.
Here, we report on our experimental investigation 
of the sequence of gluing bifurcations in an analog electronic circuit.

\section{Theoretical predictions}
\subsection{General considerations}
In the phase space of a continuous dissipative dynamical system,
the basic ingredient of the gluing bifurcation 
is an equilibrium of the saddle type with one-dimensional unstable manifold. 
We start from the situation when the system,
like the Lorenz equations, possesses a symmetry 
which transforms into each other two components of this manifold. 
When the parameters of the system are varied, location of
invariant manifolds in the phase space varies as well. At certain
combinations of parameters, one of the components of the unstable manifold 
can return back to the saddle along the stable manifold and form
the homoclinic orbit. 
Symmetry ensures that the second component returns as well: homoclinic
orbits come in pairs. 
Sequence of events (scenario) which accompany the birth/destruction 
of homoclinic orbits, depends on the leading eigenvalues of linearization 
of the equations at the saddle point. 
Since unstable manifold is one-dimensional, 
there is just one positive eigenvalue, denoted below as $\lambda_+$. 
Below we restrict ourselves to the case when
the closest to zero negative eigenvalue $\lambda_-$ is real. 
The ``saddle index'' $\nu=|\lambda_-|\lambda_+$
indicates which of the two properties -- contraction or expansion -- 
dominates the phase space in the neighborhood of the fixed point,
and, thereby, governs the stability of the bifurcating solutions.  
In absence of symmetry, destruction of a single homoclinic trajectory 
creates a unique periodic orbit which is stable if $\nu>1$ 
and unstable otherwise~\cite{Shilnikov}.  
Presence of the second, symmetric homoclinic trajectory enriches dynamics: 
all trajectories which leave the vicinity of the saddle, are re-injected back. 
In this case, under $\nu<1$ the countable set of unstable periodic orbits, 
as well as a continuum of recurrent trajectories is simultaneously born 
from the pair of homoclinic orbits~\cite{Afraimovich-Bykov-Shilnikov-77}; 
this ``homoclinic explosion'' is a crucial step 
in the subsequent formation of the Lorenz attractor. 
The situation for $\nu>1$ is simpler: 
here two stable periodic orbits approach the saddle point, 
and are ``glued together'' forming two homoclinic orbits. 
When the pair of homoclinic trajectories breaks up, 
the new stable symmetric periodic orbit is left in the phase space: 
it is obtained by concatenation of the previously existing ones.  
The length and the number of loops (turns) of the attracting trajectory 
in the phase space is doubled, like in the case of the period-doubling 
bifurcation. Notably, in contrast to the period
doubling bifurcation, the temporal period displays unbounded growth
when the system approaches the bifurcation: the period becomes 
infinite when the homoclinic trajectories are formed.

Further variation of parameter can result in the sequence 
of secondary gluing bifurcations: unstable manifold can return
to the saddle after performing several turns in the phase space.
Before this, a symmetry-breaking bifurcation should take place: 
the newborn symmetric periodic orbit is destabilized 
in the course of the pitchfork bifurcation, 
and two mutually symmetric orbits branch from it.
These two orbits approach the unstable manifold of the saddle 
and coalesce in the next gluing bifurcation. 
As a consequence, the new stable periodic orbit is born, 
which has four times more loops than the original ones. 
The subsequent scenario consists of alternating gluing- 
and symmetry-breaking bifurcations that 
eventually end in the formation of the chaotic attractor 
which has a two-lobe shape, reminiscent of the Lorenz attractor. 
The sequence of the bifurcational values of the parameter has been shown
to converge at the exponential 
rate~\cite{Arneodo-Coullet-Tresser-81a,Lyubimov-Zaks-83}.
In contrast to the period-doubling scenario, 
this rate is not the unique universal constant: 
renormalization group analysis shows that the universality class 
is completely  predetermined by the (in general, non-integer)
value of the saddle index $\nu>1$. 

Remarkably, the attractor 
which is formed in the course of this bifurcation scenario, 
occupies a certain intermediate position between order and chaos: 
the Fourier spectrum of the trajectory 
is neither discrete like in case of regular dynamics, nor absolutely 
continuous like in case of a chaotic or stochastic process, but is supported 
by the fractal set. Accordingly, observables are characterized through 
long-range non-exponential correlations~\cite{Pikovsky-Zaks-Feudel-Kurths-95}. 

If the symmetry between the components of the unstable manifold is violated, 
each gluing bifurcation is a codimension-two event. 
On the plane of two parameters, there are numerous paths which 
lead from order to chaos via the formations of secondary homoclinic orbits; 
each of these paths is characterized by its own scaling 
constants\cite{Gambaudo-Procaccia-Thomae-Tresser-86,%
Procaccia-Thomae-Tresser-87,Lyubimov-Pikovsky-Zaks-89,Zaks-93}.

\subsection{The model set of equations}
While choosing a dynamical system which should serve as a candidate 
for experimental modeling by an electronic circuit, it is reasonable
to minimize both the order of the system and the complexity of its
linear and nonlinear terms.
A unique gluing bifurcation can take place on the phase plane, but
for occurrence of {\em sequences} of such bifurcations the phase space
should be at least three-dimensional. A natural candidate is a system
of the Lorenz equations~\cite{Lorenz-63}: they possess the desired symmetry 
and are capable to exhibit homoclinic bifurcations.
In the three-dimensional parameter space of the Lorenz equations,
formation of two symmetric homoclinic orbits has codimension one: 
it takes place upon the two-dimensional surface. 
In the phase space, a homoclinic bifurcation is a nonlocal event. For this
reason, the values of parameters under which the homoclinic orbits
in dissipative dynamical systems are formed, typically 
cannot be explicitly obtained in the closed form and should be found
numerically. There are several analytical techniques which allow
to exclude presence of homoclinic orbits and/or provide bounds for their 
existence in the parameter space~\cite{Sanchez,Yu_2006,Leonov_12,Leonov_13}.
We are, however, not aware of rigorous theoretical results which would allow to 
conclude on the location of the whole bifurcation surface in the
parameter space of the Lorenz equations. 
Numerical evidence indicates that the two-dimensional surface of homoclinic 
bifurcation lies in the part of the parameter space 
which corresponds to $\nu<1$; 
hence, the ``canonical'' Lorenz equations display no gluing bifurcations.
To circumvent this obstacle, we introduce an additional term 
and consider the set of equations
\begin{eqnarray}\label{lz}
\dot{x}&=&\sigma(y-x)+A\,y\,z\nonumber\\
\dot{y}&=&R\,x -\,y\, -\,x\,z\\
\dot{z}&=&x\,y\,-b\,z\nonumber
\end{eqnarray}
Here, $\sigma$, $R$ and $b$ are the conventional Lorenz parameters, whereas 
$A$ parameterizes the added nonlinear term in the first equation. 
At $A$=0 Eqs (\ref{lz}) turn into the Lorenz equations.
The system (\ref{lz}) is reminiscent of the model, 
employed in \cite{Lyubimov-Zaks-83} for studies of thermal convection 
in the layer of fluid subjected to high-frequency modulation of gravity.
We fix the traditional values of the parameters $\sigma$=10 and 
$b$=8/3~\cite{Lorenz-63}, and vary the remaining parameters $R$ and $A$.

Among the properties of equations (\ref{lz}) we list only those 
which are relevant for gluing bifurcations and for experimental modeling.
Equations (\ref{lz}) are invariant with respect to the transformation 
$\{x\to -x,\;y\to -y\}$.  The origin $x=y=z=0$ is the equilibrium 
which is stable for $R<1$ and is a saddle with one-dimensional unstable
manifold in the parameter range $R>1$. At $R=1$ the pitchfork bifurcation 
takes place at the origin; this bifurcation is supercritical for $A<\sigma$ 
and subcritical otherwise\footnote{In the latter case, the pitchfork
is preceded by the saddle-node bifurcation which occurs at 
$\displaystyle R=2\sqrt{\frac{\sigma}{A}}-\frac{\sigma}{A}$.}.
Like in the original Lorenz equations, variation of parameters can
produce pairs of structurally unstable homoclinic orbits to the saddle point; 
these orbits leave the vicinity of the origin along the $xy$-plane 
and return to it along the $z$-axis;
under  $\sigma=10, b=8/3, A=0$ this happens at $R=13.926\ldots$.
The saddle index of the origin equals
\begin{equation}
\nu=\frac{2\,b}{-\sigma-1+\sqrt{(\sigma-1)^2+4\sigma\,R}}.
\label{saddle_index}
\end{equation}
Since the value of $\nu$ is $A$-independent,  
variation of $A$ allows to study the effects 
of additional nonlinearity under constant saddle index. 
The value of $\nu$ is smaller than 1 for $R>R_\nu=(b+\sigma)(b+1)/\sigma$ 
and exceeds 1 otherwise.
Numerical integration (we have used the recurrent Taylor algorithm
of the 30th order with variable time-step) indicates 
that increase of $A$ lowers the critical value $R_{\rm hom}$, 
required for the formation of the pair of homoclinic orbits. 
Therefore, at small positive values of $A$ breakup of homoclinic orbits
is followed by the homoclinic explosion 
and the Lorenz scenario of transition to chaos, 
whereas the sufficiently large values of $A$ ensure the inequality $\nu>1$, 
so that the sequence of gluing bifurcations is observed.

\section{Experimental setup}
\begin{figure}[h]
\includegraphics[clip,width=1\columnwidth]{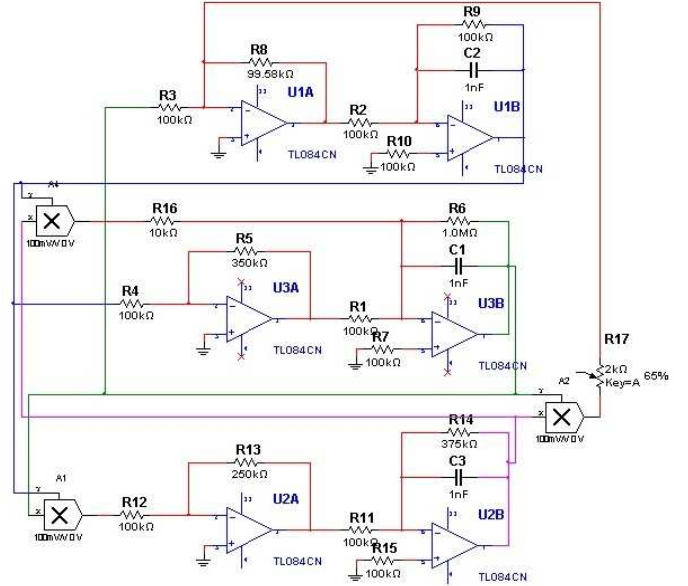} 
\caption{Basic electronic circuit which models Eqs (\ref{lz}). 
Nominal values of resistors and capacitors are shown on the scheme. 
The analog multipliers (denoted by crosses) are of the type AD633AN.
Variables $u,v,w$ are output voltage values of operational amplifiers
U1B, U2B and U3B, respectively.
}
\label{fig_1} 
\end{figure}

For our measurements we took the electronic circuit with the help of which 
the authors of~\cite{Sanchez-Matias-98} reproduced the dynamics 
on the Lorenz attractor.  In order to take account 
of the additional nonlinear term in the first equation of~(\ref{lz}),
we introduced an additional analog multiplier
(the rightmost multiplier in Fig.~\ref{fig_1}).

The cited values of electrical characteristics correspond to fixed values 
$\sigma=10$ and $b=8/3$ in Eqs (\ref{lz}). 
Changing of parameters $A$ and $R$ is enabled by variation of 
respective resistances R17 and R5. The value of $A$ equals 10
divided by the value of R17 in $k\Omega$; 
the value of $R$ equals 0.01 of the resistance R5 in $k\Omega$.

To ensure that all voltages in the circuit stay inside the operating range 
of dynamical multipliers (from --10 to 10 V), 
the original dependent and independent variables need to be rescaled: 
we assume that the dimensionless potential differences $x$, $y$ and $z$ 
are measured in volts, and proceed to $u=x/5$, $ v=y/5$,  $w=z/10$.
For the time measured in ``seconds'',  we introduce $\tau=t/T$ with $T$=100. 
This recasts Eqs (\ref{lz}) into
\begin{eqnarray}\label{rescaled} 
u'&=&T\left(\sigma(v-u)+10\,A\,v\,w\right)\nonumber\\
v'&=&T\,(R\,u -\,v\, -\,10\,u\,w)\\
w'&=&T\,(2.5\,u\,v\,-b\,w)\nonumber
\end{eqnarray}
where prime denotes differentiation with respect to $\tau$. 
Finally, it should be  noted that, 
due to inevitable presence of nonlinearities in the multipliers, 
the real circuit, in fact, does not reproduce the perfect 
symmetry of Eq.(\ref{rescaled})
with respect to the simultaneous change of sign of voltages $u$ and $v$.

In order to initiate the trajectories close to the unstable manifold
of the saddle point, the circuit was grounded for a short time. 
The subsequent growth of voltage began from rather small values.

Voltages were measured and recorded at a rate of 5000 records per second. 
We did not apply any kind of filtering; stability of the scheme and smallness 
of time-step allowed us to plot quite smooth phase portraits of the system 
directly from the experimental data. 

\section{Results}
\subsection{Lorenz scenario of transition to chaos}
We measured and recorded 
time-dependent voltages in the circuit 
for various values of the parameters $A$ and $R$
at fixed $\sigma$=10 and $b$=8/3. The equilibrium $u$=$v$=$w$=0 is stable 
for $R<$1 and is a saddle-point in the range $R>$1. 
Measurements on the circuit and numerical integration of Eqs (\ref{lz})
confirm existence of the curve $R_{\rm hom}(A)$ 
upon which principal homoclinic orbits  to this saddle are formed. 
Both in the experiment and in numerical studies  this event is marked
by change in the asymptotics of large $t$ for trajectories with
initial conditions near the origin. Prior to the bifurcation, the trajectory
which starts with small positive values of $u$ and $v$ tends to the attractor 
(equilibrium or oscillatory state) which is contained in the half-space $u>0$. 
After the bifurcation, this trajectory makes a loop in the half-space $u>0$, 
crosses into the opposite half-space $v<0$ and settles on the attractor there.

\begin{figure}[h]
\begin{centering}
\includegraphics[clip,width=1\columnwidth]{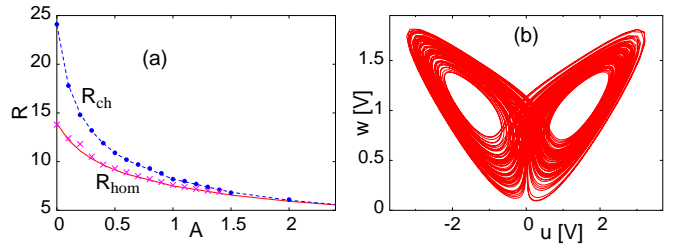} 
\par\end{centering}
\caption{Lorenz-like dynamics at $\nu<1$ 
with $\sigma$=10 and $b$=8/3.\protect\\
(a): Bifurcations on the parameter plane. 
Solid line $R_{\rm hom}$ (numerics), and crosses (experiment): 
homoclinic explosion.
Dashed line $R_{\rm ch}$ (numerics) and circles (experiment): 
onset of chaotic motion.
(b): Projection of the phase portrait at $A$=1.5, $R$=3.6 (experiment).}
\label{fig_2} 
\end{figure}

Under low values of $A$, formation of homoclinic orbit occurs at $\nu<1$;
this corresponds to the Lorenz scenario of transition to chaos. 
Immediately beyond this curve the chaotic set, created by the homoclinic
explosion at  $R_{\rm hom}(A)$, is unstable; its presence is reflected
in the existence of chaotic transients which precede relaxation to stable
equilibria (``metastable chaos''~\cite{Yorke_79}). On the parameter plane,
the domain corresponding to this kind of dynamics is bounded from above
by the curve $R_{\rm ch}(A)$ which marks stabilization of the chaotic set
and emergence of the chaotic attractor. Upon this curve the unstable manifold
of the saddle point hits the stable manifold of the unstable periodic 
orbit~\cite{Afraimovich-Bykov-Shilnikov-77}.
Numerically, we detected the latter curve by tracking the separatrices
of the saddle  point; in the experiments  $R_{\rm ch}(A)$ 
was identified as the lowest value of $R$ (or the leftmost value of $A$) 
where  the circuit possessed non-decaying chaotic dynamics.
For values of $R$ beyond $R_{\rm ch}(A)$, 
we observed irregular oscillations of voltages, with phase portraits 
strongly reminiscent of the Lorenz attractor (Fig.~\ref{fig_2}).

\subsection{Sequences of gluing bifurcations}
According to the above reasoning, homoclinic explosion should be replaced 
by the gluing bifurcation when the corresponding value of $R_{\rm hom}$ 
gets below $R_\nu$=209/45=4.6444$\ldots$; 
this corresponds to the range $A\geq$ 4. Indeed, we observed in the experiment 
the Lorenz-like chaotic attractors for $A<4$ and gluing bifurcations for $A>4$.

We illustrate the transformation of the phase portrait 
along the gluing bifurcation scenario 
with the plots of projections which correspond to increase of $A$ 
at constant values of other parameters.
\begin{figure}[h]
\includegraphics[clip,width=1\columnwidth]{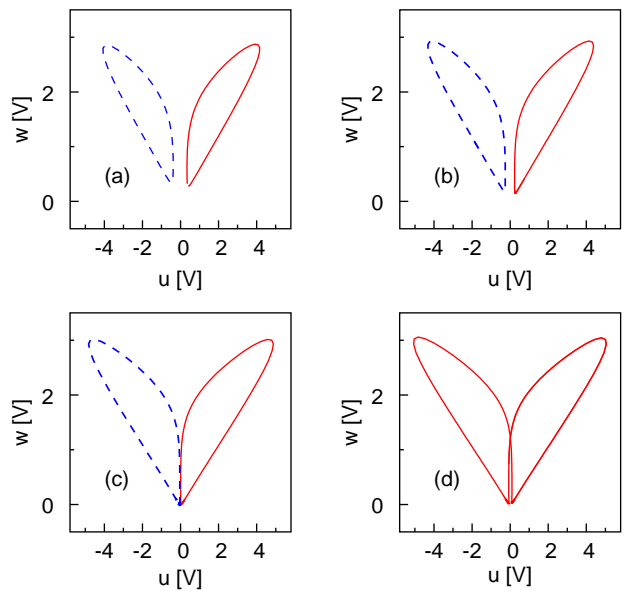} 
\caption{Principal gluing bifurcation in the circuit: 
Evolution of the attracting set. 
Parameter values: $\sigma$=10, $b$=8/3, $R$=3.6.\protect\\
(a): $A$=6.4; two stable periodic orbits.
(b): $A$=6.55; two stable periodic orbits.
(c): $A$=7.; formation of two trajectories, homoclinic to the equilibrium.
(d): $A$=7.05; stable self-symmetric orbit with two loops.}
\label{fig_3} 
\end{figure}

By varying initial conditions, 
we are able to identify in the phase space of the circuit 
two attracting closed trajectories which are almost symmetric to each other
(Fig.\ref{fig_3}a). As the parameter $A$ is increased, 
these two orbits come closer (Fig.\ref{fig_3}b), approach the invariant 
manifolds of the saddle equilibrium 
at the origin and form the pair of symmetric homoclinic orbits to this saddle
(Fig.\ref{fig_3}c). A slight increase of the parameter $A$ makes 
the homoclinic orbits break up and disappear: the only attractor of the system,
plotted in Fig.\ref{fig_3}d, is the periodic orbit with
two loops which is (up to instrumental resolution)
invariant under the symmetry transformation 
$\{u\to -u$, $v\to -v\}$ .  
Close to the homoclinic bifurcation, the measured values of
the period $T_0$ of the oscillations match the known logarithmic asymptotics: 
$T_0(A) \sim -\log(A_{\rm hom}-A)$. 

\begin{figure}[h]
\includegraphics[clip,width=0.9\columnwidth]{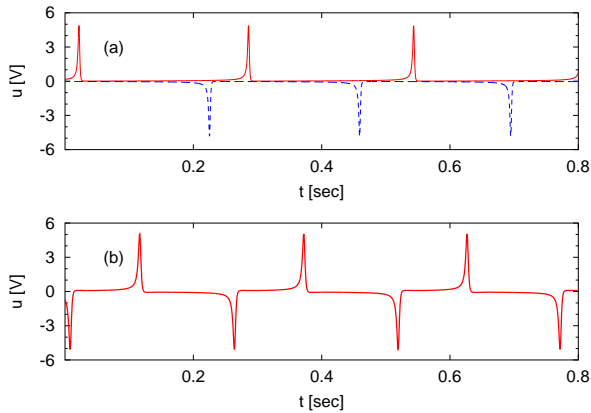} 
\caption{Temporal evolution close to the gluing bifurcation.\protect\\
(a): $A$=7.0; oscillations on limit cycles with 1 loop prior to the bifurcation.
(b): $A$=7.05; oscillations on the limit cycle with 2 loops. 
Other parameters like in Fig.~\ref{fig_3}.}
\label{fig_4} 
\end{figure}

As seen in Fig.\ref{fig_4}, the oscillations are strongly anharmonic: the
overwhelming part of the period is spent in nearly motionless state. 
This corresponds to long hovering of the trajectory 
in the vicinity of the saddle point. 

\begin{figure}[h]
\includegraphics[clip,width=.9\columnwidth]{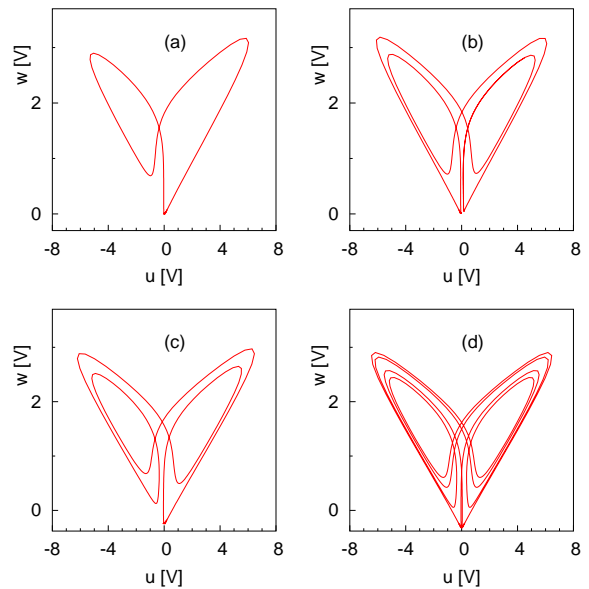} 
\caption{Further stages of the gluing scenario
(parameters: $\sigma$=10, $b$=8/3, $R$=3.6).
(a): $A$=8.05; asymmetric stable periodic orbit with two turns
(b): $A$=8.10; symmetric stable periodic orbit with 4 turns, 
born from the secondary gluing bifurcation.
(c): $A$=9.37; asymmetric stable periodic orbit with 4 turns;
(d): $A$=9.38; symmetric stable periodic orbit with 8 turns, 
born from the third gluing bifurcation.}
\label{fig_5} 
\end{figure}

Further stages of the evolution of the attracting trajectory are sketched in 
Fig.\ref{fig_5}. The symmetric limit cycle with two turns (Fig.\ref{fig_3}d)
loses stability as a result of the pitchfork bifurcation. 
One of the two resulting stable asymmetric limit cycles 
is shown in Fig.\ref{fig_5}a. As the parameter is further increased, 
these two limit cycles approach the invariant manifolds of the equilibrium, 
and the secondary gluing bifurcation takes place: two homoclinic orbits 
with 2 turns are formed. Their subsequent breakup produces 
a single symmetric stable orbit with 4 turns (Fig.\ref{fig_5}b).
Further, the events are repeated on the new level:
the symmetric orbit with 4 turns is destabilized,
and two stable asymmetric ones are born (Fig.\ref{fig_5}c). 
These asymmetric periodic orbits
merge in the next gluing bifurcation, form a pair of homoclinic
orbits with 4 turns, and leave the stable periodic orbit with 8 turns,
shown in Fig.\ref{fig_5}d.

\begin{figure}[h]
\begin{centering}
\includegraphics[clip,width=1\columnwidth]{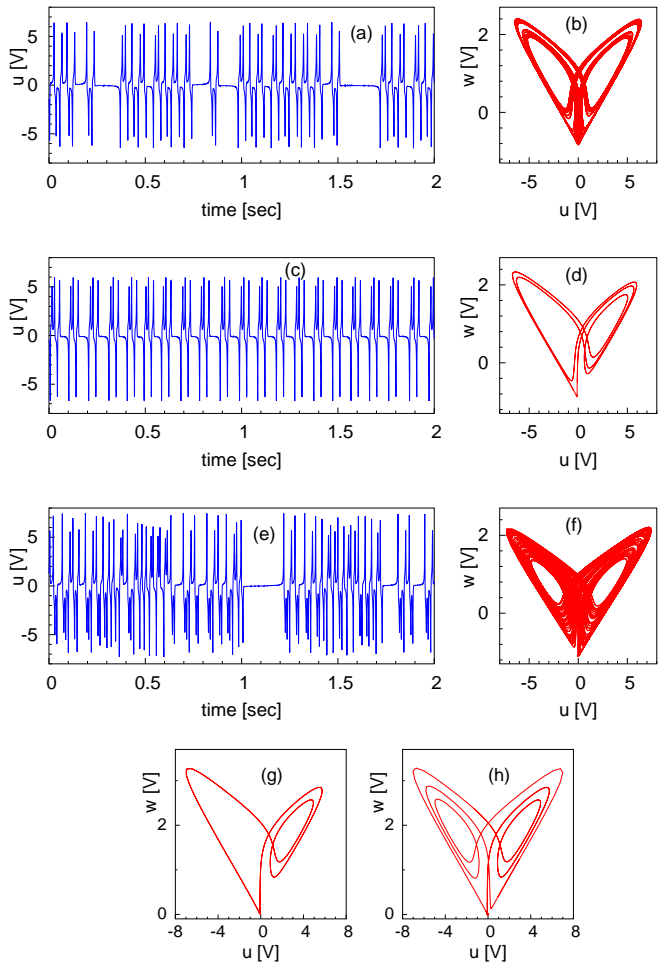} 
\par\end{centering}
\caption{Experimental records of dynamics beyond the first gluing sequence.
(a,b): Chaotic attractor at $A$=9.39.
(c,d): Limit cycle with 5 turns at $A$=10.17.
(e,f): Chaotic attractor at $A$=13.46.
(g,h) gluing bifurcation, mediated by homoclinic orbits with 3 turns 
at $A$=11.51. 
Other parameters: $\sigma$=10, $b$=8/3, $R$=3.6.}
\label{fig_6} 
\end{figure}

In the perfectly symmetric setup this bifurcation sequence would
continue, each time doubling the number of turns of the
attracting periodic orbit in the phase space~\cite{Arneodo-Coullet-Tresser-81a}.
Presence of asymmetry is known to disable the complete sequence 
of homoclinic ``doublings''~\cite{constraint,Zaks-93a}; 
mathematically, this owes to the fact,
that linearization of the corresponding renormalization operator 
near its fixed point possesses an additional
unstable direction, responsible for the asymmetry.
Finite resolution of our measurements (finite step-size for variable 
resistance R17  in Fig.~\ref{fig_1})
as well as the inevitable asymmetry in the electronic circuit 
did not allow us to resolve the further stages of the gluing process: 
instead we observe rapid emergence of the chaotic attractor  
(top row of Fig.~\ref{fig_6}).

A characteristic feature of chaotic oscillations in this circuit,
well recognizable in the oscillograms of Fig.~\ref{fig_6}(a,e),
is  presence of relatively long nearly quiescent plateaus: 
they correspond to hovering near the saddle point. Further increase
of the parameter $A$ has disclosed that the range in which
oscillations are chaotic,
is interspersed by numerous windows in which various stable symmetric
and asymmetric periodic oscillations are observed. One of such 
periodic states is shown in Fig.~\ref{fig_6}(c,d). In terms of the
parameter $A$, 
periodic windows are bounded from above
by secondary homoclinic bifurcations and serve as starting states 
of further gluing scenarios. An example of bifurcation, in which two periodic
regimes with three turns glue together and create a periodic orbit with
six turns is shown in Fig.~\ref{fig_6}(g,h). It should be noted, that
increase of the resistance R17 enhances asymmetry 
in the circuit, therefore, in general, homoclinic trajectories formed,
respectively, by the ``left'' and the ``right'' components of the unstable
manifold occur at slightly different values of the parameter $A$.    

\section{Discussion and Outlook}

We have demonstrated in the experiment on the analog electronic circuit
the Lorenz scenario of transition to chaos as well as
several initial stages of the sequence of gluing bifurcations.  
The circuit mimics the modified Lorenz system, and we utilized
the fact that modification allows, by variation of a single resistance,
to move across the parameter space without changing the value
of the saddle index. Observed sequence of transformations
of phase portraits provides an unambiguous qualitative 
confirmation of theoretical predictions. For a quantitative comparison
(convergence rate of the bifurcation scenario and other scaling constants), 
the experimental resolution -- in particular the step-size of the active
parameter -- should be improved.

In a wider context of dynamical systems with gluing bifurcations, 
it should be noted that a straightforward coding procedure
maps temporal evolution of voltages onto  the binary alphabet: 
a symbol 1 is assigned to each turn of the orbit in the half-space $u>0$,
and a symbol 0 marks each turn in the half-space $u<0$.  
In this sense, a pair of binary codes which 
correspond to two components of the unstable manifold of the saddle, provides
a complete characterization of dynamics~\cite{Procaccia-Thomae-Tresser-87}.
In presence of symmetry, one code is sufficient; remarkably,
it is related to the kneading sequence of the family of logistic mappings.
In case of noticeable asymmetry, both kneading sequences are needed; 
here, certain aspects of the description are analogous to the formalism 
for (in general, discontinuous) circle mappings: rotation numbers, 
devil staircase etc.~\cite{Zaks-93}. 
Since our experimental results match well the theoretical description
of the route to chaos for symmetric systems, we may hope that 
explicit introduction of controlled asymmetry into equations (\ref{rescaled}) 
and into the analog circuit of Fig.~\ref{fig_1}
should allow to detect in the circuit bifurcation also those scenarios
which theory predicts for asymmetric gluing bifurcations.

\end{document}